\documentclass[aps,preprint,nofootinbib,floatfix]{revtex4}%
\usepackage{amssymb}
\usepackage{amsmath}
\usepackage{epsfig}
\usepackage{amsfonts}
\usepackage{graphicx}%
\begin{document}
\title{High-energy String Scatterings of Compactified Open String}
\author{Jen-Chi Lee}
\email{jcclee@cc.nctu.edu.tw}
\affiliation{Department of Electrophysics, National Chiao-Tung University and Physics
Division, National Center for Theoretical Sciences, Hsinchu, Taiwan, R.O.C.}
\author{Tomohisa Takimi}
\email{tomotakimi@mail.nctu.edu.tw}
\affiliation{Department of Electrophysics, National Chiao-Tung University and Physics
Division, National Center for Theoretical Sciences, Hsinchu, Taiwan, R.O.C.}
\author{Yi Yang}
\email{yyang@phys.cts.nthu.edu.tw}
\affiliation{Department of Electrophysics, National Chiao-Tung University and Physics
Division, National Center for Theoretical Sciences, Hsinchu, Taiwan, R.O.C.}

\begin{abstract}
We calculate high-energy massive string scattering amplitudes of compactified
open string. We derive infinite linear relations, or stringy symmetries, among
soft high-energy string scattering amplitudes of different string states in
the Gross kinematic regime (GR). In addition, we systematically analyze all
hard power-law and soft exponential fall-off regimes of high-energy
compactified open string scatterings by comparing the scatterings with their
26D noncompactified counterparts. In particular, we discover the existence of
a power-law regime at fixed angle and an exponential fall-off regime at small
angle for high-energy compactified open string scatterings. The linear
relations break down as expected in all power-law regimes. The analysis can be
extended to the high-energy scatterings of the compactified closed string,
which corrects and extends the previous results in \cite{Compact}.

\end{abstract}
\maketitle

\section{Introduction}

There are three fundamental characteristics of high-energy, fixed angle string
scattering amplitudes \cite{GM, Gross, GrossManes}, which are not shared by
the field theory scatterings. These are the softer exponential fall-off
behavior \cite{Veneziano,GM} (in contrast to the hard power-law behavior of
field theory scatterings \cite{BF}), the infinite Regge-pole structure
\cite{Closed,LY} of the form factor and the existence of infinite number of
linear relations \cite{ChanLee1,ChanLee2,
CHL,CHLTY,PRL,paperB,susy,Closed,Dscatt,Decay,HL,LY}, or stringy symmetries,
discovered recently among string scattering amplitudes of different string
states in the high-energy string scattering amplitudes. An important new
ingredient to derive these linear relations is the zero-norm states (ZNS)
\cite{ZNS1,ZNS3,ZNS2} in the old covariant first quantized (OCFQ) string
spectrum. Other approaches related to this development can be found in
\cite{West,Rey,Moore,MO,Regge, Soldate}

It was believed \cite{Wall,Compact} that the newly discovered linear relations
are responsible for the softer exponential fall-off behavior of high-energy
string scatterings. One way to justify this conjecture is to find more hard
power-law high-energy string scatterings, which simultaneously show the
breakdown of the above linear relations. However, it is well known that the
genetic high-energy, fixed angle behavior of string scatterings is soft
exponential fall-off rather than hard power-law. Recently, following an old
suggestion of Mende \cite{Mende}, two of the present authors \cite{Compact}
calculated high-energy massive scattering amplitudes of closed bosonic string
with some coordinates compactified on the torus. They obtained infinite linear
relations among high-energy scattering amplitudes of different string states
in the Gross kinematic regime (GR). This result is reminiscent of the
existence of an infinite number of massive ZNS in the compactified closed
string spectrum constructed in \cite{Lee}. In addition, they discovered that,
for some kinematic regime, these infinite linear relations break down and,
simultaneously, the string amplitudes enhance to hard power-law behavior
instead of the usual soft exponential fall-off behavior at high energies.

To further understand the relationship of the infinite linear relations and
the softer exponential fall-off behavior of high-energy, fixed angle string
scatterings, it is crucial to find more examples of high-energy string
scatterings, which show the unusual power-law behavior and, simultaneously,
give the breakdown of the infinite linear relations. In this paper, we
calculate high-energy massive string scattering amplitudes of open bosonic
string with some coordinates compactified on the torus. As in the case of
compactified closed string, we obtain infinite linear relations among
soft\textit{\ }high-energy scattering amplitudes of different string states in
the GR. This result is reminiscent of the existence of an infinite number of
massive ZNS in the compactified open string spectrum constructed in
\cite{Lee2}. More importantly, we analyze all possible hard power-law and soft
exponential fall-off regimes of high-energy compactified open string
scatterings by comparing the scatterings with their 26D noncompactified
counterparts. In particular, we discover the existence of a power-law regime
at fixed angle and an exponential fall-off regime at small angle for
high-energy compactified open string scatterings. These new phenomena never
happen in the 26D string scatterings. The linear relations break down as
expected in all power-law regimes. The analysis can be extended to the
high-energy scatterings of the compactified closed string, which corrects and
extends the previous results in \cite{Compact}. In particular, we correct the
"Mende regime" discussed in \cite{Compact}, which is indeed exponential
fall-off behaved rather than power-law claimed in \cite{Compact}. As an
example, we derive a hard power-law regime at fixed angle for high-energy
compactified closed string scatterings. This paper is organized as following.
In section II we calculated high-energy massive scattering amplitudes of
compactified open string. In section III we classify all kinematic regimes of
the amplitudes and extend our results to the closed string case. A brief
conclusion is given in section IV.

\section{High-energy Scatterings}

We consider 26D open bosonic string with one coordinate compactified on
$S^{1}$ with radius $R$. As we will see later, it is straightforward to
generalize our calculation to more compactified coordinates. The mode
expansion of the compactified coordinate is%
\begin{equation}
X^{25}\left(  \sigma,\tau\right)  =x^{25}+K^{25}\tau+i\sum_{k\neq0}%
\frac{\alpha_{k}^{25}}{k}e^{-ik\tau}\cos n\sigma
\end{equation}
where $K^{25}$ is the canonical momentum in the $X^{25}$ direction%
\begin{equation}
K^{25}=\frac{2\pi l-\theta_{j}+\theta_{i}}{2\pi R}.
\end{equation}
Note that $l$ is the quantized momentum and we have included a nontrivial
Wilson line with $U(n)$ Chan-Paton factors, $i,j=1,2...n.$, which will be
important in the later discussion. The mass spectrum of the theory is
\begin{equation}
M^{2}=\left(  K^{25}\right)  ^{2}+2\left(  N-1\right)  \equiv\left(
\frac{2\pi l-\theta_{j}+\theta_{i}}{2\pi R}\right)  ^{2}+\hat{M}^{2}
\label{mass}%
\end{equation}
where we have defined level mass as $\hat{M}^{2}=2\left(  N-1\right)  $ and
$N=\sum_{k\neq0}\alpha_{-k}^{25}\alpha_{k}^{25}+\alpha_{-k}^{\mu}\alpha
_{k}^{\mu},\mu=0,1,2...24.$ We are going to consider 4-point correlation
function in this paper. In the center of momentum frame, the kinematic can be
set up to be \cite{Compact}%

\begin{align}
k_{1}  &  =\left(  +\sqrt{p^{2}+M_{1}^{2}},-p,0,-K_{1}^{25}\right)  ,\\
k_{2}  &  =\left(  +\sqrt{p^{2}+M_{2}^{2}},+p,0,+K_{2}^{25}\right)  ,\\
k_{3}  &  =\left(  -\sqrt{q^{2}+M_{3}^{2}},-q\cos\phi,-q\sin\phi,-K_{3}%
^{25}\right)  ,\\
k_{4}  &  =\left(  -\sqrt{q^{2}+M_{4}^{2}},+q\cos\phi,+q\sin\phi,+K_{4}%
^{25}\right)
\end{align}
where $p$ is the incoming momentum, $q$ is the outgoing momentum and $\phi$ is
the center of momentum scattering angle. In the high-energy limit, one
includes only momenta on the scattering plane, and we have included the fourth
component for the compactified direction as the internal momentum. The
conservation of the fourth component of the momenta implies%
\begin{equation}
K_{1}^{25}-K_{2}^{25}+K_{3}^{25}-K_{4}^{25}=0.
\end{equation}
Note that%
\begin{equation}
k_{i}^{2}=K_{i}^{2}-M_{i}^{2}=-\hat{M}_{i}^{2}.
\end{equation}
The center of mass energy $E$ is defined as (for large $p,q$)%
\begin{equation}
E=\dfrac{1}{2}\left(  \sqrt{p^{2}+M_{1}^{2}}+\sqrt{p^{2}+M_{2}^{2}}\right)
=\dfrac{1}{2}\left(  \sqrt{q^{2}+M_{3}^{2}}+\sqrt{q^{2}+M_{4}^{2}}\right)  .
\end{equation}
We have%
\begin{align}
-k_{1}\cdot k_{2}  &  =\sqrt{p^{2}+M_{1}^{2}}\cdot\sqrt{p^{2}+M_{2}^{2}}%
+p^{2}+K_{1}^{25}K_{2}^{25}\nonumber\\
&  =\dfrac{1}{2}\left(  s+k_{1}^{2}+k_{2}^{2}\right)  =\dfrac{1}{2}s-\frac
{1}{2}\left(  \hat{M}_{1}^{2}+\hat{M}_{2}^{2}\right)  ,\label{k1k2}\\
-k_{2}\cdot k_{3}  &  =-\sqrt{p^{2}+M_{2}^{2}}\cdot\sqrt{q^{2}+M_{3}^{2}%
}+pq\cos\phi+K_{2}^{25}K_{3}^{25}\nonumber\\
&  =\dfrac{1}{2}\left(  t+k_{2}^{2}+k_{3}^{2}\right)  =\dfrac{1}{2}t-\frac
{1}{2}\left(  \hat{M}_{2}^{2}+\hat{M}_{3}^{2}\right)  ,\label{t}\\
-k_{1}\cdot k_{3}  &  =-\sqrt{p^{2}+M_{1}^{2}}\cdot\sqrt{q^{2}+M_{3}^{2}%
}-pq\cos\phi-K_{1}^{25}K_{3}^{25}\nonumber\\
&  =\dfrac{1}{2}\left(  u+k_{1}^{2}+k_{3}^{2}\right)  =\dfrac{1}{2}u-\frac
{1}{2}\left(  \hat{M}_{1}^{2}+\hat{M}_{3}^{2}\right)  \label{u}%
\end{align}
where $s,t$ and $u$ are the Mandelstam variables with%
\begin{equation}
s+t+u=\sum_{i}\hat{M}_{i}^{2}\sim2\left(  N-4\right)  .
\end{equation}
Note that the Mandelstam variables defined above are not the usual
$25$-dimensional Mandelstam variables in the scattering process since we have
included the internal momentum $K_{i}^{25}$ in the definition of $k_{i}$. We
are now ready to calculate the high-energy scattering amplitudes. In the
high-energy limit, we define the polarizations on the scattering plane to be%

\begin{align}
e^{P}  &  =\frac{1}{M_{2}}\left(  \sqrt{p^{2}+M_{2}^{2}},p,0,0\right)  ,\\
e^{L}  &  =\frac{1}{M_{2}}\left(  p,\sqrt{p^{2}+M_{2}^{2}},0,0\right)  ,\\
e^{T}  &  =\left(  0,0,1,0\right)
\end{align}
where the fourth component refers to the compactified direction. It is easy to
calculate the following relations%
\begin{align}
e^{P}\cdot k_{1}  &  =-\frac{1}{M_{2}}\left(  \sqrt{p^{2}+M_{1}^{2}}%
\sqrt{p^{2}+M_{2}^{2}}+p^{2}\right)  ,\label{MR}\\
e^{P}\cdot k_{3}  &  =\frac{1}{M_{2}}\left(  \sqrt{q^{2}+M_{3}^{2}}\sqrt
{p^{2}+M_{2}^{2}}-pq\cos\phi\right)  ,
\end{align}%
\begin{align}
e^{L}\cdot k_{1}  &  =-\frac{p}{M_{2}}\left(  \sqrt{p^{2}+M_{1}^{2}}%
+\sqrt{p^{2}+M_{2}^{2}}\right)  ,\\
e^{L}\cdot k_{3}  &  =\frac{1}{M_{2}}\left(  p\sqrt{q^{2}+M_{3}^{2}}%
-q\sqrt{p^{2}+M_{2}^{2}}\cos\phi\right)  , \label{GR}%
\end{align}%
\begin{equation}
e^{T}\cdot k_{1}=0\text{, \ \ }e^{T}\cdot k_{3}=-q\sin\phi.
\end{equation}

In this paper, we will consider the case of a tensor state \cite{Compact}%

\begin{equation}
\left(  \alpha_{-1}^{T}\right)  ^{N-2r}\left(  \alpha_{-2}^{L}\right)
^{r}\left\vert k_{2},l_{2},i,j\right\rangle
\end{equation}
at a general mass level $\hat{M}_{2}^{2}=2\left(  N-1\right)  $ scattered with
three "tachyon" states (with $\hat{M}_{1}^{2}=\hat{M}_{3}^{2}=\hat{M}_{4}%
^{2}=-2$). In general, we could have considered the more general high-energy state%

\begin{equation}
\left(  \alpha_{-1}^{T}\right)  ^{N-2r-2m-\sum_{n}ns_{n}}\left(  \alpha
_{-1}^{L}\right)  ^{2m}\left(  \alpha_{-2}^{L}\right)  ^{r}\prod
\limits_{n}\left(  \alpha_{-n}^{25}\right)  ^{s_{n}}\left\vert k_{2}%
,l_{2},i,j\right\rangle .
\end{equation}
However, for our purpose here and for simplicity, we will not consider the
general vertex in this paper. The $s-t$ channel of the high-energy scattering
amplitude can be calculated to be (We will ignore the trace factor due to
Chan-Paton in the scattering amplitude calculation . This does not affect our
final results in this paper)
\begin{align}
A &  =\int d^{4}x\left\langle e^{ik_{1}X}\left(  x_{1}\right)  \left(
\partial X^{T}\right)  ^{N-2r}\left(  i\partial^{2}X^{L}\right)  ^{r}%
e^{ik_{2}X}\left(  x_{2}\right)  e^{ik_{3}X}\left(  x_{3}\right)  e^{ik_{4}%
X}\left(  x_{4}\right)  \right\rangle \nonumber\\
&  =\int d^{4}x\cdot\prod\limits_{i<j}\left(  x_{i}-x_{j}\right)  ^{k_{i}\cdot
k_{j}}\nonumber\\
&  \cdot\left[  \frac{ie^{T}\cdot k_{1}}{x_{1}-x_{2}}+\frac{ie^{T}\cdot k_{3}%
}{x_{3}-x_{2}}+\frac{ie^{T}\cdot k_{4}}{x_{4}-x_{2}}\right]  ^{N-2r}%
\cdot\left[  \frac{e^{L}\cdot k_{1}}{\left(  x_{1}-x_{2}\right)  ^{2}}%
+\frac{e^{L}\cdot k_{3}}{\left(  x_{3}-x_{2}\right)  ^{2}}+\frac{e^{L}\cdot
k_{4}}{\left(  x_{4}-x_{2}\right)  ^{2}}\right]  ^{r}.
\end{align}
After fixing the $SL(2,R)$ gauge and using the kinematic relations derived
previously, we have%
\begin{align}
A &  =i^{N}\left(  -1\right)  ^{k_{1}\cdot k_{2}+k_{1}\cdot k_{3}+k_{2}\cdot
k_{3}}\left(  q\sin\phi\right)  ^{N-2r}\left(  \frac{1}{M_{2}}\right)
^{r}\cdot\int_{0}^{1}dx\cdot x^{k_{1}\cdot k_{2}}\left(  1-x\right)
^{k_{2}\cdot k_{3}}\cdot\left[  \frac{1}{1-x}\right]  ^{N-2r}\nonumber\\
&  \cdot\left[  \frac{p\left(  \sqrt{p^{2}+M_{1}^{2}}+\sqrt{p^{2}+M_{2}^{2}%
}\right)  }{x^{2}}-\frac{\left(  p\sqrt{q^{2}+M_{3}^{2}}-q\sqrt{p^{2}%
+M_{2}^{2}}\cos\phi\right)  }{\left(  1-x\right)  ^{2}}\right]  ^{r}%
\nonumber\\
&  =\left(  -1\right)  ^{k_{1}\cdot k_{2}+k_{1}\cdot k_{3}+k_{2}\cdot k_{3}%
}\left(  iq\sin\phi\right)  ^{N}\left(  -\frac{\left(  p\sqrt{q^{2}+M_{3}^{2}%
}-q\sqrt{p^{2}+M_{2}^{2}}\cos\phi\right)  }{M_{2}q^{2}\sin^{2}\phi}\right)
^{r}\nonumber\\
&  \cdot\sum_{i=0}^{r}\binom{r}{i}\left[  -\frac{p\left(  \sqrt{p^{2}%
+M_{1}^{2}}+\sqrt{p^{2}+M_{2}^{2}}\right)  }{\left(  p\sqrt{q^{2}+M_{3}^{2}%
}-q\sqrt{p^{2}+M_{2}^{2}}\cos\phi\right)  }\right]  ^{i}\cdot\int_{0}%
^{1}dx\cdot x^{k_{1}\cdot k_{2}-2i}\left(  1-x\right)  ^{k_{2}\cdot
k_{3}-N+2i}\nonumber\\
&  =\left(  -iq\sin\phi\right)  ^{N}\left(  -\frac{\left(  p\sqrt{q^{2}%
+M_{3}^{2}}-q\sqrt{p^{2}+M_{2}^{2}}\cos\phi\right)  }{M_{2}q^{2}\sin^{2}\phi
}\right)  ^{r}\nonumber\\
&  \cdot\sum_{i=0}^{r}\binom{r}{i}\left[  -\frac{p\left(  \sqrt{p^{2}%
+M_{1}^{2}}+\sqrt{p^{2}+M_{2}^{2}}\right)  }{\left(  p\sqrt{q^{2}+M_{3}^{2}%
}-q\sqrt{p^{2}+M_{2}^{2}}\cos\phi\right)  }\right]  ^{i}\cdot B\left(
-\frac{1}{2}s+N-2i-1,-\frac{1}{2}t+2i-1\right)
\end{align}
where $B(u,v)$ is the Euler beta function. We can do the high-energy
approximation of the gamma function $\Gamma\left(  x\right)  $ then do the
summation, and end up with
\begin{align}
A &  =\left(  -iq\sin\phi\right)  ^{N}\left(  -\frac{\left(  p\sqrt
{q^{2}+M_{3}^{2}}-q\sqrt{p^{2}+M_{2}^{2}}\cos\phi\right)  }{M_{2}q^{2}\sin
^{2}\phi}\right)  ^{r}\nonumber\\
&  \cdot\sum_{i=0}^{r}\binom{r}{i}\left[  -\frac{p\left(  \sqrt{p^{2}%
+M_{1}^{2}}+\sqrt{p^{2}+M_{2}^{2}}\right)  }{\left(  p\sqrt{q^{2}+M_{3}^{2}%
}-q\sqrt{p^{2}+M_{2}^{2}}\cos\phi\right)  }\right]  ^{i}\cdot\frac
{\Gamma\left(  -1-\frac{1}{2}s+N-2i\right)  \Gamma\left(  -1-\frac{1}%
{2}t+2i\right)  }{\Gamma\left(  2+\frac{1}{2}u\right)  }\nonumber\\
&  \simeq\left(  -iq\sin\phi\right)  ^{N}\left(  -\frac{\left(  p\sqrt
{q^{2}+M_{3}^{2}}-q\sqrt{p^{2}+M_{2}^{2}}\cos\phi\right)  }{M_{2}q^{2}\sin
^{2}\phi}\right)  ^{r}\nonumber\\
&  \cdot\sum_{i=0}^{r}\binom{r}{i}\left[  -\frac{p\left(  \sqrt{p^{2}%
+M_{1}^{2}}+\sqrt{p^{2}+M_{2}^{2}}\right)  }{\left(  p\sqrt{q^{2}+M_{3}^{2}%
}-q\sqrt{p^{2}+M_{2}^{2}}\cos\phi\right)  }\right]  ^{i}\nonumber\\
&  \cdot B\left(  -1-\dfrac{1}{2}s,-1-\frac{1}{2}t\right)  \left(
-1-\dfrac{1}{2}s\right)  _{N-2i}\left(  -1-\frac{1}{2}t\right)  _{2i}\left(
2+\dfrac{1}{2}u\right)  _{-N}\label{power}%
\end{align}
where $(a)_{j}=a(a+1)(a+2)...(a+j-1)$ is the Pochhammer symbol, and
$(a)_{j}=(a)^{j}$ for large $a$ and fixed $j.$

\section{\bigskip Classification of Compactified String Scatterings}

It is well known that there are two kinematic regimes for the high-energy
string scatterings in 26D open bosonic string theory. The UV behavior of the
finite and fixed angle scatterings in the GR is soft exponential fall-off.
Moreover, there exist infinite linear relations among scatterings of different
string states in this regime \cite{ChanLee1,ChanLee2,
CHL,CHLTY,PRL,paperB,susy,Closed,Dscatt,Decay}. On the other hand, the UV
behavior of the small angle scatterings in the Regge regime is hard power-law.
The linear relations break down in the Regge regime. As we will see soon, the
UV structure of the compactified open string scatterings is more richer. In
the following, we will systematically analyze all possible hard power-law
regimes of high-energy compactified open string scatterings by comparing the
scatterings with their noncompactified counterparts. In particular, we show
that all hard power-law regimes of high-energy compactified open string
scatterings can be traced back to the Regge regime of the 26D high-energy
string scatterings. The linear relations break down as expected in all
power-law regimes. The analysis can be extended to the high-energy scatterings
of the compactified closed string, which corrects and extends the previous
results in \cite{Compact}.

\subsection{\bigskip Gross Regime - Linear Relations}

In the Gross regime, $p^{2}\simeq q^{2}\gg K_{i}^{2}$ and $p^{2}\simeq
q^{2}\gg N$, Eq.(\ref{power}) reduces to%
\begin{equation}
A\simeq\left(  -iE\frac{\sin\frac{\phi}{2}}{\cos\frac{\phi}{2}}\right)
^{N}\left(  -\frac{1}{2M_{2}}\right)  ^{r}\cdot B\left(  -1-\frac{1}%
{2}s,-1-\frac{1}{2}t\right)  . \label{beta}%
\end{equation}
For each fixed mass level $N$, we have the linear relation for the scattering
amplitudes%
\begin{equation}
\frac{\mathcal{T}^{\left(  n,r\right)  }}{\mathcal{T}^{\left(  n,0\right)  }%
}=\left(  -\frac{1}{2M_{2}}\right)  ^{r} \label{linear}%
\end{equation}
with coefficients consistent with our previous results
\cite{ChanLee1,ChanLee2, CHL,CHLTY,PRL,paperB,susy,Closed,Dscatt,Decay}. Note
that in Eq.(\ref{beta}) there is an exponential fall-off factor in the
high-energy expansion of the beta function. The infinite linear relation in
Eq.(\ref{linear}) "soften" the high-energy behavior of string scatterings in
the GR.

\subsection{\bigskip Classification of compactified open string}

We first discuss the open string case. Since our definitions of the Mandelstam
variables $s,t$ and $u$ in Eq.(\ref{k1k2}) to Eq.(\ref{u}) include the
compactified coordinates, we can analyze the UV structure of the compactified
string scatterings by comparing the scatterings with their simpler 26D
counterparts. We introduce the space part of the momentum vectors%

\begin{align}
\mathbf{k}_{1}  &  =\left(  -p,0,-K_{1}^{25}\right)  ,\\
\mathbf{k}_{2}  &  =\left(  +p,0,+K_{2}^{25}\right)  ,\\
\mathbf{k}_{3}  &  =\left(  -q\cos\phi,-q\sin\phi,-K_{3}^{25}\right)  ,\\
\mathbf{k}_{4}  &  =\left(  +q\cos\phi,+q\sin\phi,+K_{4}^{25}\right)  ,
\end{align}
and define the "26D scattering angle" $\widetilde{\phi}$ as following%
\begin{equation}
\mathbf{k}_{1}\cdot\mathbf{k}_{3}=\left\vert \mathbf{k}_{1}\right\vert
\left\vert \mathbf{k}_{3}\right\vert \cos\widetilde{\phi}. \label{26D}%
\end{equation}
It is then easy to see that the UV behavior of the compactified string
scatterings is power-law if and only if $\widetilde{\phi}$ is small. This
criterion can be used to classify all possible power-law and exponential
fall-off kinematic regimes of high-energy compactified open string
scatterings. We first consider the high-energy scatterings with one coordinate compactified.

\subsubsection{\bigskip Compactified 25D scatterings}

$\boldsymbol{I.}$ For the case of $\phi=$ finite, the only choice to achieve
UV power-law behavior is to require (we choose $K_{1}^{25}\simeq K_{2}%
^{25}\simeq K_{3}^{25}\simeq K_{4}^{25}$ and $p\simeq q$ in the following
discussion)%
\begin{equation}
\left(  K_{i}^{25}\right)  ^{2}\gg p^{2}\simeq q^{2}\gg N. \label{newpower}%
\end{equation}
By the criterion of Eq.(\ref{26D}), this is a power-law regime. To explicitly
show that this choice of kinematic regime does lead to UV power-law behavior,
we will show that it implies
\begin{equation}
s=\text{ constant} \label{const}%
\end{equation}
in the open string scattering amplitudes, which in turn gives the desire
power-law behavior of high-energy compactified open string scattering in
Eq.(\ref{power}). On the other hand, it can be shown that the linear relations
break down as expected in this regime. For the choice of kinematic regime in
Eq.(\ref{newpower}) , Eq.(\ref{k1k2}) and Eq.(\ref{const}) imply
\begin{equation}
\lim_{p\rightarrow\infty}\frac{\sqrt{p^{2}+M_{1}^{2}}\cdot\sqrt{p^{2}%
+M_{2}^{2}}+p^{2}}{K_{1}^{25}K_{2}^{25}}=\lim_{p\rightarrow\infty}\frac
{\sqrt{p^{2}+M_{1}^{2}}\cdot\sqrt{p^{2}+M_{2}^{2}}+p^{2}}{\left(  \frac{2\pi
l_{1}-\theta_{j,1}+\theta_{i,1}}{2\pi R}\right)  \left(  \frac{2\pi
l_{2}-\theta_{j,2}+\theta_{i,2}}{2\pi R}\right)  }=-1. \label{condition}%
\end{equation}
For finite momenta $l_{1}$and $l_{2}$, Eq.(\ref{condition}) can be achieved by
scattering of string states with "super-highly" winding nontrivial Wilson
lines%
\begin{equation}
(\theta_{i,1}-\theta_{j,1})\rightarrow\infty\text{, \ }(\theta_{i,2}%
-\theta_{j,2})\rightarrow-\infty. \label{wilson}%
\end{equation}
A careful analysis for this choice gives%
\begin{equation}
(\lambda_{1}+\lambda_{2})^{2}=0 \label{lamda}%
\end{equation}
where signs of $\lambda_{1}=\frac{p}{K_{1}^{25}}$ and $\lambda_{2}=-\frac
{p}{K_{2}^{25}}$ are chosen to be the same. It can be seen now that the
kinematic regime in Eq.(\ref{newpower}) does solve Eq.(\ref{lamda}).

We now consider the second possible regime for the case of $\phi=$ finite,
namely%
\begin{equation}
\left(  K_{i}^{25}\right)  ^{2}\simeq p^{2}\simeq q^{2}\gg N. \label{oldmende}%
\end{equation}
By the criterion of Eq.(\ref{26D}), this is an exponential fall-off regime. To
explicitly show that this choice of kinematic regime does lead to UV
exponential fall-off behavior, we see that, for this regime, it is impossible
to achieve Eq.(\ref{const}) since Eq.(\ref{lamda}) has no nontrivial solution.
Note that there are no linear relations in this regime. Although
$\widetilde{\phi}=$ finite in this regime, it is different from the GR in the
26D scatterings since $K_{i}^{25}$ is as big as the scattering energy $p.$
\textit{In conclusion, we have discovered a }$\phi=$ \textit{finite regime}
\textit{with} \textit{UV power-law behavior for the high-energy compactified
open string scatterings. }This new phenomenon never happens in the 26D string
scatterings. \textit{The linear relations break down as expected in this
regime.}

$\bigskip\boldsymbol{II.}$ For the case of small angle $\phi\simeq0$
scattering, we consider the first power-law regime%
\begin{equation}
qK_{1}^{25}=-pK_{3}^{25}\text{ and }\left(  K_{i}^{25}\right)  ^{2}\simeq
p^{2}\simeq q^{2}\gg N. \label{uconst}%
\end{equation}
By the criterion of Eq.(\ref{26D}), this is a power-law regime. To explicitly
show that this choice of kinematic regime does lead to UV power-law behavior,
we will show that it implies
\begin{equation}
u=\text{ constant} \label{uconstant}%
\end{equation}
in the open string scattering amplitudes, which in turn gives the desire
power-law behavior of high-energy compactified open string scattering in
Eq.(\ref{power}). For this choice of kinematic regime, Eq.(\ref{u}) and
Eq.(\ref{uconst}) imply
\begin{equation}
\lim_{p\rightarrow\infty}\frac{\sqrt{p^{2}+M_{1}^{2}}\cdot\sqrt{q^{2}%
+M_{3}^{2}}+pq}{K_{1}^{25}K_{3}^{25}}=-1. \label{ratio}%
\end{equation}
By choosing different sign for $K_{1}^{25}$ and $K_{3}^{25}$, Eq.(\ref{ratio})
can be solved for any real number $\lambda\equiv\frac{p}{K_{1}^{25}}=-\frac
{q}{K_{3}^{25}}.$

The second choice for the power-law regime is the same as Eq.(\ref{newpower})
in the $\phi=$ finite regime. The proof to show that it is indeed a power-law
regime is similar to the proof in section $I.$

The last choice for the power-law regime is%
\begin{equation}
\left(  K_{i}^{25}\right)  ^{2}\ll p^{2}\simeq q^{2}\gg N.
\end{equation}
It is easy to show that this is indeed a power-law regime.

The last kinematic regime for the case of small angle $\phi\simeq0$ scattering
is%
\begin{equation}
qK_{1}^{25}\neq-pK_{3}^{25}\text{ and }\left(  K_{i}^{25}\right)  ^{2}\simeq
p^{2}\simeq q^{2}\gg N.
\end{equation}
By the criterion of Eq.(\ref{26D}), this is an exponential fall-off regime. We
give one example here. Let's choose $\lambda\equiv\frac{p}{K_{1}^{25}}%
\neq-\frac{q}{K_{3}^{25}}=2\lambda.$ By choosing different sign for
$K_{1}^{25}$ and $K_{3}^{25}$, Eq.(\ref{ratio}) reduces to%
\begin{equation}
\lambda^{2}=0,
\end{equation}
which has no nontrivial solution for $\lambda$, and one can not achieve the
power-law condition Eq.(\ref{uconstant}). So this is an exponential fall-off
regime. \textit{In conclusion, we have discovered a }$\phi\simeq0$ regime
\textit{with} \textit{UV exponential fall-off behavior for the high-energy
compactified open string scatterings. }This new phenomenon never happens in
the 26D string scatterings. This completes the classification of all kinematic
regimes for compactified 25D scatterings.

\subsubsection{\bigskip Compactified 24D (or less) scatterings}

For this case, we need to introduce another parameter to classify the UV
behavior of high-energy scatterings, namely the angle $\delta$ between
$\vec{K}_{1}$ and $\vec{K}_{2}$, $\vec{K}_{1}\cdot\vec{K}_{2}=\left\vert
K_{1}\right\vert \left\vert K_{2}\right\vert \cos\delta.$ Similar results can
be easily derived through the same method used in the compactified 25D
scatterings. The classification is independent of the details of the moduli
space of the compact spaces. We summarize the results in the following table:

\begin{center}%
\begin{tabular}
[c]{|c|c|c|c|c|}\hline
$\phi$ & $\tilde{\phi}$ & UV Behavior & Examples of the Kinematic Regimes &
Linear Relations\\\hline
&  &  & \multicolumn{1}{|l|}{$\vec{K}_{i}^{2}\ll p^{2}\simeq q^{2}\gg N$} &
Yes\\\cline{4-5}%
finite & finite & Exponential fall-off & \multicolumn{1}{|l|}{$\vec{K}_{i}%
^{2}\simeq p^{2}\simeq q^{2}\gg N$} & \\\cline{4-4}
&  &  & \multicolumn{1}{|l|}{$\vec{K}_{i}^{2}\gg p^{2}\simeq q^{2}\gg N$ and
$\cos\delta\neq0$} & No\\\cline{2-4}
& $\tilde{\phi}\simeq0$ & Power-law & \multicolumn{1}{|l|}{$\vec{K}_{i}^{2}\gg
p^{2}\simeq q^{2}\gg N$ and $\cos\delta=0$} & \\\hline
&  &  & \multicolumn{1}{|l|}{$\vec{K}_{i}^{2}\ll p^{2}\simeq q^{2}\gg N$} &
\\\cline{4-4}
& $\tilde{\phi}\simeq0$ & Power-law & \multicolumn{1}{|l|}{$\vec{K}_{i}%
^{2}\simeq p^{2}\simeq q^{2}\gg N$ and $q\vec{K}_{1}=-p\vec{K}_{3}$} &
\\\cline{4-4}%
$\phi\simeq0$ &  &  & \multicolumn{1}{|l|}{$\vec{K}_{i}^{2}\gg p^{2}\simeq
q^{2}\gg N$ and $\cos\delta=0$} & No\\\cline{2-4}
& finite & Exponential fall-off & \multicolumn{1}{|l|}{$\vec{K}_{i}^{2}\simeq
p^{2}\simeq q^{2}\gg N$ and $q\vec{K}_{1}\neq-p\vec{K}_{3}$} & \\\cline{4-4}
&  &  & \multicolumn{1}{|l|}{$\vec{K}_{i}^{2}\gg p^{2}\simeq q^{2}\gg N$ and
$\cos\delta\neq0$} & \\\hline
\end{tabular}

\end{center}

\subsection{\bigskip Classification of compactified closed string}

The classification for the high-energy compactified open string scatterings
can be easily extended to the case of compactified closed string
\cite{Compact}. For illustration, we discuss the case of $\phi=$ finite regime
with UV power-law behavior for the high-energy 25D compactified closed string
scatterings. The kinematic set up for the compactified closed string are
\begin{align}
k_{1L,R}  &  =\left(  +\sqrt{p^{2}+M_{1}^{2}},-p,0,-K_{1L,R}\right)  ,\\
k_{2L,R}  &  =\left(  +\sqrt{p^{2}+M_{2}^{2}},+p,0,+K_{2L,R}\right)  ,\\
k_{3L,R}  &  =\left(  -\sqrt{q^{2}+M_{3}^{2}},-q\cos\phi,-q\sin\phi
,-K_{3L,R}\right)  ,\\
k_{4L,R}  &  =\left(  -\sqrt{q^{2}+M_{4}^{2}},+q\cos\phi,+q\sin\phi
,+K_{4L,R}\right)
\end{align}
where the left and right compactified momenta are defined to be%
\begin{equation}
K_{L,R}=K\pm L=\frac{m}{R}\pm\dfrac{1}{2}nR
\end{equation}
where $m$ is the quantized momentum and $n$ is the winding number. The
"super-highly" winding nontrivial Wilson line for the open string in
Eq.(\ref{wilson}) is replaced by "super-highly" closed string winding $n_{i}.$
The condition to achieve power-law behavior for the compactified open string
scatterings, Eq.(\ref{const}), is replaced by
\begin{equation}
s_{L}=\text{ constant, }s_{R}=\text{ constant } \label{closedpower}%
\end{equation}
for the compactified closed string scatterings. The left and the right
Mandelstam variables are defined to be%
\begin{align}
s_{L,R}  &  \equiv-(k_{1L,R}+k_{2L,R})^{2},\\
t_{L,R}  &  \equiv-(k_{2L,R}+k_{3L,R})^{2},\\
u_{L,R}  &  \equiv-(k_{1L,R}+k_{3L,R})^{2},
\end{align}
with%
\begin{equation}
s_{L,R}+t_{L,R}+u_{L,R}=\sum_{i}M_{iL,R}^{2}. \label{sum}%
\end{equation}
One can easily calculate the following kinematic relations%
\begin{align}
-k_{1L,R}\cdot k_{2L,R}  &  =\sqrt{p^{2}+M_{1}^{2}}\cdot\sqrt{p^{2}+M_{2}^{2}%
}+p^{2}+K_{1L,R}K_{2L,R}\nonumber\\
&  =\dfrac{1}{2}\left(  s_{L,R}+k_{1L,R}^{2}+k_{2L,R}^{2}\right)  =\dfrac
{1}{2}s_{L,R}-\frac{1}{2}\left(  M_{1L,R}^{2}+M_{2L,R}^{2}\right)  ,
\label{k1*k2}%
\end{align}%
\begin{align}
-k_{2L,R}\cdot k_{3L,R}  &  =-\sqrt{p^{2}+M_{2}^{2}}\cdot\sqrt{q^{2}+M_{3}%
^{2}}+pq\cos\phi+K_{2L,R}K_{3L,R}\nonumber\\
&  =\dfrac{1}{2}\left(  t_{L,R}+k_{2L,R}^{2}+k_{3L,R}^{2}\right)  =\dfrac
{1}{2}t_{L,R}-\frac{1}{2}\left(  M_{2L,R}^{2}+M_{3L,R}^{2}\right)  ,
\end{align}%
\begin{align}
-k_{1L,R}\cdot k_{3L,R}  &  =-\sqrt{p^{2}+M_{1}^{2}}\cdot\sqrt{q^{2}+M_{3}%
^{2}}-pq\cos\phi-K_{1L,R}K_{3L,R}\nonumber\\
&  =\dfrac{1}{2}\left(  u_{L,R}+k_{1L,R}^{2}+k_{3L,R}^{2}\right)  =\dfrac
{1}{2}u_{L,R}-\frac{1}{2}\left(  M_{1L,R}^{2}+M_{3L,R}^{2}\right)  .
\label{newLTY}%
\end{align}
It was calculated in \cite{Compact} that the beta function part of the
high-energy closed string scattering amplitude can be written as%
\begin{align}
&  A_{closed}\sim B\left(  -1-\dfrac{t_{R}}{2},-1-\dfrac{u_{R}}{2}\right)
B\left(  -1-\dfrac{t_{L}}{2},-1-\dfrac{u_{L}}{2}\right) \nonumber\\
&  =\frac{\Gamma\left(  -\frac{t_{R}}{2}-1\right)  \Gamma\left(  -\frac{u_{R}%
}{2}-1\right)  }{\Gamma\left(  \frac{s_{R}}{2}+2\right)  }\frac{\Gamma\left(
-\frac{t_{L}}{2}-1\right)  \Gamma\left(  -\frac{u_{L}}{2}-1\right)  }%
{\Gamma(\frac{s_{L}}{2}+2)}.
\end{align}
It is easy to see that the condition, Eq.(\ref{closedpower}), leads to the
power-law behavior of the compactified closed string scattering amplitudes. To
satisfy the condition in Eq.(\ref{closedpower}), we define the following
"super-highly" winding kinematic regime%
\begin{equation}
n_{i}^{2}\gg p^{2}\simeq q^{2}\gg N_{R}+N_{L}. \label{closedregime}%
\end{equation}
Note that all $m_{i}$ were chosen to vanish in order to satisfy the
conservations of compactified momentum and winding number respectively
\cite{Compact}. For the choice of the kinematic regime in
Eq.(\ref{closedregime}), Eq.(\ref{k1*k2}) and Eq.(\ref{closedpower}) imply%
\begin{equation}
\lim_{p\rightarrow\infty}\frac{\sqrt{p^{2}+M_{1}^{2}}\cdot\sqrt{p^{2}%
+M_{2}^{2}}+p^{2}}{2(K_{1}K_{2}+L_{1}L_{2})}=\lim_{p\rightarrow\infty}%
\frac{\sqrt{p^{2}+M_{1}^{2}}\cdot\sqrt{p^{2}+M_{2}^{2}}+p^{2}}{2\left(
\dfrac{m_{1}m_{2}}{R^{2}}+\dfrac{1}{4}n_{1}n_{2}R^{2}\right)  }=-1.
\label{exist}%
\end{equation}
Note that since we have set $m_{i}=0,$ Eq.(\ref{exist}) is similar to
Eq.(\ref{condition}) for the compactified open string case, and one can get
nontrivial solution for Eq.(\ref{lamda}) with signs of $\lambda_{1}=\frac
{2p}{n_{1}R}$ and $\lambda_{2}=-\frac{2p}{n_{2}R}$ the same. This completes
the discussion of power-law regime at fixed angle for high-energy compactified
closed string scatterings. The "super-highly" winding regime derived in this
subsection is to correct the "Mende regime"
\begin{equation}
E^{2}\simeq M^{2}\gg N_{R}+N_{L} \label{wrong}%
\end{equation}
discussed in \cite{Compact}. The regime defined in Eq.(\ref{wrong}) is indeed
exponential fall-off behaved rather than power-law claimed in \cite{Compact}.

\section{Conclusion}

In this paper, we calculate high-energy massive string scattering amplitudes
of compactified open string. We derive infinite linear relations among soft
high-energy string scattering amplitudes of different string states at
arbitrary but fixed mass level in the Gross kinematic regime (GR). We then
systematically analyze all hard power-law and soft exponential fall-off
regimes of high-energy compactified open string scatterings by comparing the
scatterings with their 26D noncompactified counterparts. We classify all
kinematic regimes for high-energy compatified open string scatterings. In
particular, we discover the existence of a power-law regime at fixed angle and
an exponential fall-off regime at small angle for high-energy compactified
open string scatterings. These two new phenomena do not exist for high-energy
26D noncompactified open string scatterings. It is a stringy effect due to
string compactification. The linear relations break down as expected in all
power-law regimes. Finally, we have extended the analysis to the high-energy
scatterings of the compactified closed string, which corrects and extends the
previous results in \cite{Compact}.

\section{Acknowledgments}

This work is supported in part by the National Science Council, 50 Billions
Project of MOE and National Center for Theoretical Science, Taiwan, R.O.C.


\begin{thebibliography}{99}                                                                                               %


\bibitem {GM}D.~J.~Gross and P.~F.~Mende,
%``The High-Energy Behavior Of String Scattering Amplitudes,''
Phys.\ Lett.\ B \textbf{197}, 129 (1987);
%``String Theory Beyond The Planck Scale,''
Nucl.\ Phys.\ B \textbf{303}, 407 (1988).

\bibitem {Gross}D.~J.~Gross,
%``High-Energy Symmetries Of String Theory,''
Phys.\ Rev.\ Lett.\ \textbf{60}, 1229 (1988); Phil.\ Trans.\ R. Soc. Lond.
A329, 401 (1989).

\bibitem {GrossManes}D.~J.~Gross and J.~L.~Manes,
%``The High-Energy Behavior Of Open String Scattering,''
Nucl.\ Phys.\ B \textbf{326}, 73 (1989). See section 6 for details.

\bibitem {Veneziano}G. Veneziano, Nuovo Cimento A57 (1968)190.

\bibitem {BF}S.J. Brodsky and G.R. Farrar, Phys. Rev. D11,1309 (1975); C.G.
Callan Jr. and D.J. Gross, Phys. Rev. D11,2905 (1975).

\bibitem {Closed}C.~T.~Chan, J.~C.~Lee and Y.~Yang,
%``High energy scattering amplitudes of superstring theory,''
Nucl.\ Phys.\ B \textbf{749}, 280 (2006).
%[arXiv:hep-th/0510247].
%%CITATION = HEP-TH 0510247;%%


\bibitem {LY}J.C. Lee and Y. Yang, Nucl. Phys. B798, 198 (2008).

\bibitem {ChanLee1}C.~T.~Chan and J.~C.~Lee,
%``Stringy symmetries and their high-energy limits,''
Phys.\ Lett.\ B \textbf{611}, 193 (2005).
%[arXiv:hep-th/0312226].
J.~C.~Lee,
%``Stringy symmetries and their high-energy limit''.
[arXiv:hep-th/0303012].

\bibitem {ChanLee2}C.~T.~Chan and J.~C.~Lee,
%``Zero-norm states and high-energy symmetries of string theory,''
Nucl.\ Phys.\ B \textbf{690}, 3 (2004).
%[arXiv:hep-th/0401133].


\bibitem {CHL}C.~T.~Chan, P.~M.~Ho and J.~C.~Lee,
%``Ward identities and high-energy scattering amplitudes in string theory,''
Nucl.\ Phys.\ B \textbf{708}, 99 (2005).
%[arXiv:hep-th/0410194].


\bibitem {CHLTY}C.~T.~Chan, P.~M.~Ho, J.~C.~Lee, S.~Teraguchi and Y.~Yang,
%``Solving all 4-point correlation functions for bosonic open string theory in
%the high energy limit,''
Nucl.\ Phys.\ B \textbf{725}, 352 (2005).
%[arXiv:hep-th/0504138].


\bibitem {PRL}C.~T.~Chan, P.~M.~Ho, J.~C.~Lee, S.~Teraguchi and Y.~Yang,
%``Solving all 4-point correlation functions for bosonic open string theory in
%the high energy limit,''
Phys. Rev. Lett. 96 (2006) 171601.

\bibitem {paperB}C.~T.~Chan, P.~M.~Ho, J.~C.~Lee, S.~Teraguchi and Y.~Yang,
%``High energy scattering amplitudes of superstring theory,''
Nucl.\ Phys.\ B \textbf{749}, 266 (2006).
%[arXiv:hep-th/0510247].
%%CITATION = HEP-TH 0510247;%%


\bibitem {susy}C.~T.~Chan, J.~C.~Lee and Y.~Yang,
%``High energy scattering amplitudes of superstring theory,''
Nucl.\ Phys.\ B \textbf{738}, 93 (2006).
%[arXiv:hep-th/0510247].
%%CITATION = HEP-TH 0510247;%%


\bibitem {Dscatt}C.~T.~Chan, J.~C.~Lee and Y.~Yang, " Scatterings of massive
string states from D-brane and their linear relations at high energies",
Nucl.Phys.B\textbf{764}, 1 (2007).
%[arXiv:hep-th/0212196].


\bibitem {Decay}J.C. Lee and Y. Yang, "Linear Relations of High Energy
Absorption/Emission Amplitudes of D-brane", Phys.Lett. B646 (2007) 120, hep-th/0612059.

\bibitem {HL}Pei-Ming Ho, Xue-Yan Lin, Phys.Rev. D73 (2006) 126007.

\bibitem {ZNS1}J.~C.~Lee,
%``New Symmetries Of Higher Spin States In String Theory,''
Phys.\ Lett.\ B \textbf{241}, 336 (1990); Phys.\ Rev.\ Lett.\ \textbf{64},
1636 (1990). J.~C.~Lee and B.~Ovrut,
%``Zero Norm States And Enlarged Gauge Symmetries Of Closed Bosonic
%String In Background Massive Fields,''
Nucl.\ Phys.\ B \textbf{336}, 222 (1990); J.C.Lee, Phys.\ Lett.\ B
\textbf{326}, 79 (1994).
%[arXiv:hep-th/0503005].


\bibitem {ZNS3}T.~D.~Chung and J.~C.~Lee,
%``Discrete gauge states and W(infinity) charges in c = 1 2-d gravity,''
Phys.\ Lett.\ B \textbf{350}, 22 (1995).
%[arXiv:hep-th/9412095];
%``Superfield form of discrete gauge states in c = 1 2-d supergravity,''
Z.\ Phys.\ C \textbf{75}, 555 (1997).
%[arXiv:hep-th/9505107].
J.~C.~Lee,
%``SO(2,C) invariant discrete gauge states in Liouville gravity coupled
%to minimal conformal matter,''
Eur.\ Phys.\ J.\ C \textbf{1}, 739 (1998).
%[arXiv:hep-th/0501075].


\bibitem {ZNS2}H.~C.~Kao and J.~C.~Lee,
%``Decoupling of degenerate positive-norm states in Witten's string field
%theory,''
Phys.\ Rev.\ D \textbf{67}, 086003 (2003).
%[arXiv:hep-th/0212196].
C.~T.~Chan, J.~C.~Lee and Y.~Yang,
%``Anatomy of zero-norm states in string theory,''
Phys.\ Rev.\ D \textbf{71}, 086005 (2005).
%[arXiv:hep-th/0501020].


\bibitem {West}P.~C.~West,
%``Physical states and string symmetries,''
Mod.\ Phys.\ Lett.\ A \textbf{10}, 761 (1995).
%[arXiv:hep-th/9411029].
%%CITATION = HEP-TH 9411029;%%
%\cite{Moeller:2005ez}
%\bibitem{Moeller:2005ez}
N.~Moeller and P.C.~West,
%``Arbitrary four string scattering at high energy and fixed angle,''
Nucl.\ Phys.\ B \textbf{729}, 1 (2005).
%arXiv:hep-th/0507152.
%%CITATION = HEP-TH 0507152;%%


\bibitem {Rey}J. Polchinski and M.J. Strassler, Phys.Rev.Lett. 88, 031601
(2002); JHEP 0305, 012 (2003). S.J.Rey and J.T. Yee, Nucl.Phys.B 671 (2003) 95.

\bibitem {Moore}G.~W.~Moore,
%``Finite in all directions,''
[arXiv:hep-th/9305139];
%%CITATION = HEP-TH 9305139;%%
G.~W.~Moore,
%``Symmetries of the bosonic string S matrix,''
[arXiv:hep-th/9310026].
%%CITATION = HEP-TH 9310026;%%


\bibitem {MO}P.F. Mende and H. Ooguri, Nucl. Phys. B339, 641 (1990).

\bibitem {Regge}D. Amati, M. Ciafaloni and G. Veneziano, Phys. Lett. B197, 81
(1987); Int. Jour. Mod. Phys. A3, 1615 (1988); Phys. Lett. B216, 41 (1989).

\bibitem {Soldate}M. Soldate, Phys. Lett. B186, 321, (1987); I. Muzinich and
M. Soldate, Phys. Rev. D37, 359 (1988).

\bibitem {Wall}C.~T.~Chan, J.~C.~Lee and Y.~Yang, "Power-law Behavior of
Strings Scattered from Domain-wall and Breakdown of Their High Energy Linear
Relations", hep-th/0610219.

\bibitem {Compact}J.C. Lee and Y. Yang, "Linear Relations and their Breakdown
in High Energy Massive String Scatterings in Compact Spaces", Nucl.Phys. B784
(2007) 22.
%[arXiv:hep-th/0212196].


\bibitem {Mende}Paul F. Mende, "High Energy String Collisions in a Compact
Space", Phys.Lett. B326 (1994) 216, hep-th/9401126.

\bibitem {Lee}Jen-Chi Lee, \textquotedblleft Soliton gauge states and
T-duality of closed Bosonic string compatified on torus\textquotedblright,
Eur.Phys.C7,669 (1999), hep-th/0005227.

\bibitem {Lee2}Jen-Chi Lee,"Chan-Paton soliton gauge states of the compatified
open string", Eur.Phys.C13,695 (2000), hep-th/0005228.
\end{thebibliography}
\end{document}